Persistence length and plateau modulus of semiflexible entangled polymers in primitive chain network simulations

[1,*]Yuichi Masubuchi, [2]Manlio Tassieri, [3]Giovanni Ianniruberto and [3]Giuseppe Marrucci

[1]Department of Materials Physics, Nagoya University, Nagoya 4648603, Japan

[2]Division of Biomedical Engineering, James Watt School of Engineering, Advanced Research Centre, University of Glasgow, Glasgow G11 6EW, UK

[3]Dipartimento di Ingegneria Chimica, dei Materiali e della Produzione Industriale, Università degli Studi di Napoli "Federico II", Piazzale Tecchio 80-80125 Napoli, Italy

*to whom correspondence should be addressed

mas@mp.pse.nagoya-u.ac.jp

Ver. Nov 8, 2024

ABSTRACT

The relationship between the plateau modulus ($G_0$) and the persistence length ($L_p$) of entangled semiflexible polymers is still uncertain. Some previous theoretical models have predicted that $G_0$ decreases with increasing $L_p$, while experiments indicated the opposite. In this study, we extend the primitive chain network (PCN) model to incorporate bending rigidity at the scale of the entanglement length, consistently with the coarse graining of the model. Simulations investigate the effects of rigidity (and of molecular weight) on chain conformation and relaxation modulus. Our results reveal a relationship described as $G_0 \sim L_p^{2/3}$, indicating that $G_0$ increases with increasing $L_p$, consistently with experiments. It should be considered, however, that implicit in our simulations is the condition that the entanglement length is kept fixed with changing $L_p$.

INTRODUCTION

The relation between plateau modulus $G_0$ and persistence length $L_p$ of entangled semiflexible polymers has yet to be fully elucidated[1,2]. Theoretical attempts have been made for the tightly entangled regime defined by the relation $a < L_p$, where $a$ is the entanglement mesh size in the tube model. According to the binary collision approximation by Morse[3,4], $G_0$ changes with $L_p$ as $G_0 \propto L_p^{-1/5}$. Morse[3,4] also derived another expression based on the effective medium approximation, which is $G_0 \propto L_p^{-1/3}$. Anyhow, both these expressions predict that $G_0$ decreases with increasing $L_p$, and some other theories[5–7] are also consistent with these relations. On the other hand, the theory for semiflexible gels[8] gives $G_0 \propto L_p^{7/5}$, suggesting the opposite trend. Finally, experimental data by Schuldt et al.[9] for DNA solutions indicate $G_0 \propto L_p$, while data for F-actin solutions by Tassieri et al.[10,11] even show $G_0 \propto L_p^5$.

A recent coarse grained molecular simulation[12] for crosslinked networks demonstrates that the shear modulus decreases with increasing $L_p$, and it shows an upturn at a specific critical value. If a similar result could be extended to entangled networks, the above mentioned theories and experiments might refer to different regions of $L_p$ values. However, the critical value of $L_p$ found in those simulations is much larger than that in the experiments. Moreover, the authors[12] attribute the increase of modulus to the glassy contribution, which is not significant for solutions. In conclusion, the reason for the apparent failure of theories that predict a decreasing modulus with increasing persistence length in semiflexible polymer solutions is not clear.

Here the tube model[13] is replaced by a network model inclusive of a force balance at the network nodes, with chains coarse grained at the level of the distance between consecutive entanglements. Sliding of the chains through the nodes accounts for reptation, and suitable choices of processes taking place at the chain ends provide renewal of the network topology. Chain rigidity operates between consecutive chain strands, consistently with the adopted coarse graining. As the fast-relaxing chain ends do not contribute to the modulus, a dependence of the modulus on molecular mass is expected[14,15]. Therefore, simulations for different chain lengths are necessary, and the plateau modulus definition is better related to a chain of infinite length, although obviously actual experiments cannot achieve such a limit.

In this study, we extend the primitive chain network (PCN) model[16,17] to include chain bending rigidity. We extract the relationship between plateau modulus and persistence length by performing a series of simulations with varying rigidity and molecular weight. Additionally, we obtain a relationship between chain conformation and chain rigidity. The results for the modulus suggest a new scaling relationship, $G_0 \propto L_p^{2/3}$. In the following section, we briefly formulate the simulation model. Next, we report and discuss simulation results. Finally, we draw some conclusions.

SIMULATION MODEL

As mentioned in the Introduction, simulations are based on the PCN model[16–22], where entangled polymers are replaced by a network consisting of nodes, strands, and dangling ends. Each polymer chain corresponds to a path connecting two chain ends through several nodes and strands. A slip-link is located at each node to bundle two chains, thus restricting chain motion perpendicularly to the chain contour. The equation of motion for the position of network nodes and dangling ends is written as follows:

$$0 = \mathbf{F}_d + \mathbf{F}_t + \mathbf{F}_b + \mathbf{F}_o + \mathbf{F}_R \qquad (1)$$

where $\mathbf{F}_d$ is the drag force given by:

$$\mathbf{F}_d = -\zeta \dot{\mathbf{R}} \qquad (2)$$

with $\zeta$ the node friction coefficient, and $\mathbf{R}$ its position.

The force $\mathbf{F}_t$ is the tension acting in the network strands. A key simplification is introduced here: since we focus solely on the plateau modulus and ignore predictions of terminal relaxation times (which are expected to be extremely long[4,9]), we assume the strands to be Gaussian, and composed by $n$ (virtual) Kuhn segments of length $b$. Hence, we write

$$\mathbf{F}_t = \frac{3k_B T}{b^2} \sum_i \frac{\mathbf{r}_i}{n_i} \qquad (3)$$

with $k_B T$ the thermal energy, and $\mathbf{r}_i$ the strand vector. The simplification embodied in eq 3 drastically reduces computation times.

The force related to polymer semiflexibility $\mathbf{F}_b$ (or bending force) is obtained from the potential $U_b$ as

$$\mathbf{F}_b = -\frac{\partial U_b}{\partial \mathbf{R}} \tag{4}$$

$$U_b = K \sum (\cos \theta_{i,i-1} - 1)^2 \tag{5}$$

where $K$ is the rigidity parameter, and $\theta_{i,i-1}$ is the angle between two adjacent strand vectors along the chain.

The osmotic force $\mathbf{F}_o$ is meant to suppress density fluctuations, and is related to the chemical potential $\mu$ of network strands

$$\mathbf{F}_o = -n_0 \frac{\partial \mu}{\partial \mathbf{R}} \tag{6}$$

$$\mu = \varepsilon \frac{\partial}{\partial n} \left(\frac{\phi}{\phi_0} - 1\right)^2 \tag{7}$$

where $n_0$ is the average value of $n_i$ in the simulation box, $\phi$ is the local density of Kuhn segments, $\phi_0$ its average value, and $\varepsilon$ is an energy parameter tuning compressibility, typically chosen as $\varepsilon = k_B T$.

Finally, $\mathbf{F}_R$ is the Gaussian random force obeying fluctuation-dissipation theorem:

$$\langle \mathbf{F}_R \rangle = \mathbf{0} \tag{8}$$

$$\langle \mathbf{F}_{Ri}(t) \mathbf{F}_{Rj}(t') \rangle = 2 k_B T \zeta \delta_{ij} \delta(t - t') \mathbf{I} \tag{9}$$

In addition to the evolution of $\mathbf{R}$, chain sliding through the slip-links is calculated through the 1D version of a force balance similar to eq 1, determining the rate of change of $n_i$:

$$0 = -\zeta_s \frac{\dot{n}_i}{\rho} + \frac{3 k_B T}{b^2} \left(\frac{r_i}{n_i} - \frac{r_{i-1}}{n_{i-1}}\right) + f_o + f_r \tag{10}$$

Here, the first term is the drag force, with $\zeta_s$ the sliding friction coefficient chosen as $\zeta_s = \zeta/2$ consistently with the binary assumption of entanglements, and $\rho$ the curvilinear Kuhn segment density given by $n_0/a$ (where, we recall, $a = b\sqrt{n_0}$ is the mesh size of the network). The second term is the tension difference between adjacent strands, while $f_o$ is the osmotic force along the chain generated by the chemical potential in eq 7, and $f_r$ is the Gaussian random force obeying $\langle f_r \rangle = 0$ and $\langle f_{ri}(t) f_{rj}(t') \rangle = 2 k_B T \zeta_s \delta_{ij} \delta(t - t')$.

Note that eq 10 does not contain a term linked to bending rigidity because $U_b$ does not generate any sliding force along the chain.

In addition to the aforementioned equations governing $\{\mathbf{R}_i\}$ and $\{n_i\}$, the network topology changes because of creation and destruction of entanglements at the chain ends. To account for those processes, we monitor the value of $n_i$ in each dangling end, and when $n_i$ falls below $n_0/2$ due to the chain sliding, we remove the network node to which the dangling end is connected, thereby releasing the bundled partner strands. *Vice versa*, when $n_i$ becomes larger than $3n_0/2$, we create a new network node by hooking another strand from the surrounding strands (within a distance of $a$). To avoid an anomalous increase of $U_b$ due to the creation of the new slip-link, we calculate the change $\Delta U_b$ resulting from the hooking process and accept the move with the Boltzmann probability $\exp(-\Delta U_b/k_B T)$. If the trial hooking is not accepted, we repeat the process until it is successful.

We choose $a$, $k_B T$, and $\tau = \zeta a^2/(6k_B T)$ as units of length, energy, and time, respectively, and we make all quantities nondimensional hereafter according to these units. A cubic simulation box with the box size fixed at $(12)^3$ was used, with periodic boundary conditions. The network strand density $\nu$ was fixed at 10. After a sufficiently long equilibration, stress fluctuations were recorded, and the linear relaxation modulus $G(t)$ was calculated via the Green-Kubo formula. Following earlier studies[13,14], we define the plateau modulus as $G_0 \equiv G(t = \tau)$. Eight independent simulation runs were conducted for each condition to obtain reasonable statistics.

RESULTS AND DISCUSSION

Figure 1 shows typical snapshots of the simulations. Here, the average number of strands per chain is 20. As the bending parameter $K$ increases, the chain swells as expected.

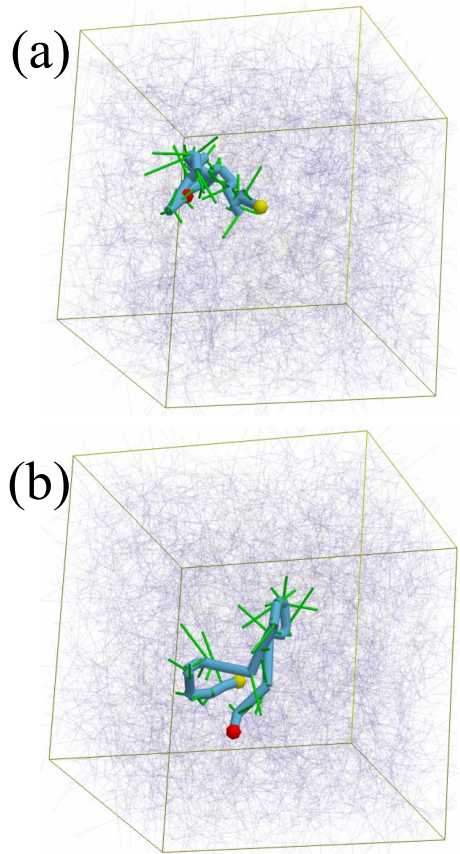

**Figure 1** Typical snapshots of the simulation box. One of the chains in the system is drawn as a series of blue cylinders, while thin violet lines show the other chains. Green bold lines are entangled partner strands of the highlighted chain. For clarity, small red and yellow balls mark the position of the chain ends. The average number $Z$ of entangled strands per chain is 20. The rigidity parameter $K$ is zero in (a), and 0.6 in (b).

Figure 2(a) shows the chain square end-to-end distance $R^2$ as a function of the average number $Z$ of entangled strands, for several values of $K$. The use of the average value of entanglements per chain is due to the fact that the number of entangled strands per chain fluctuates with time. We find that $R^2$ increases with increasing $Z$ with slope 1 (in log scales) for $K = 0$, as expected, whereas the slope somewhat increases above unity with increasing $K$. We also find that for $K$ different from zero $Z$ slightly deviates from the initially imposed value. Specifically, the average distance between consecutive entanglements (the average strand length) is somewhat sensitive to the bending force, and it slightly increases with increasing $K$.

For any assigned value of the bending parameter $K$, the values of $R^2$ and of the average primitive chain contour length $L$ obtained from the simulations enable us to derive the persistence length $L_p$ of the semiflexible chains using the following equation, originally derived for worm-like chains [23]:

$$R^2 = 2L_p^2\left[\frac{L}{L_p} - 1 + \exp\left(-\frac{L}{L_p}\right)\right] \qquad (11)$$

Figure 2(b) shows the so obtained persistence length $L_p$ as a function of $K$. Notice that, for $K = 0$ it is $L_p = 0.5a = 0.5$ (because $a = 1$), confirming that chain statistics without bending rigidity follows that of ideal chains. As expected, $L_p$ increases with increasing $K$ (up to a value of 4 when $K$ reaches 1 in Figure 2b).

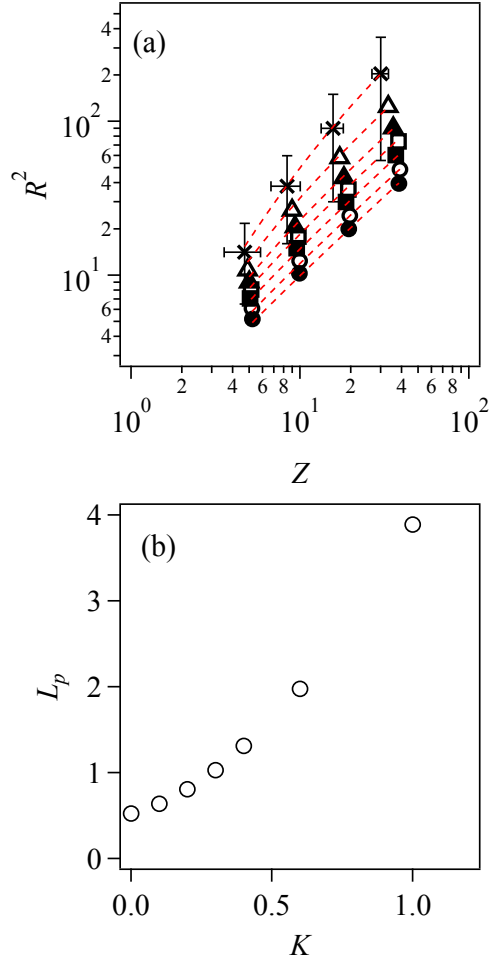

**Figure 2** (a) Chain square end-to-end distance $R^2$ as a function of the average strand number per chain $Z$ for values of the rigidity parameter $K$ equal to: 0 - filled circles, 0.1 - unfilled circles, 0.2 - filled squares, 0.3 - unfilled squares, 0.4 - filled triangles, 0.6 - unfilled triangles, 1.0 - crosses.

Error bars for $K = 1$ show the spread of $R^2$ and $Z$ values. Red dashed lines are eq 11, where (for each $Z$) $L = Za \approx Z$. (b) Persistence length $L_p$ from eq 11 as a function of $K$.

Figure 3 shows typical relaxation moduli, where the modulus is normalized by the strand density $\nu$. As previously reported, PCN simulations with $K = 0$ semi-quantitatively reproduce the dynamics of entangled flexible polymers[14,24–26]. Compared to that case, the shape of the relaxation curves $G(t)$ is not strongly altered by the bending rigidity. This insensitivity arises from the simplifying assumption that subchains (i.e., strands) are Gaussian, so the longest relaxation time remains unaffected by rigidity, as $K$ does not influence chain sliding. On the other hand, $G(t)$ shifts upwards as $K$ increases, indicating an increase of plateau modulus $G_0$ with increasing $L_p$.

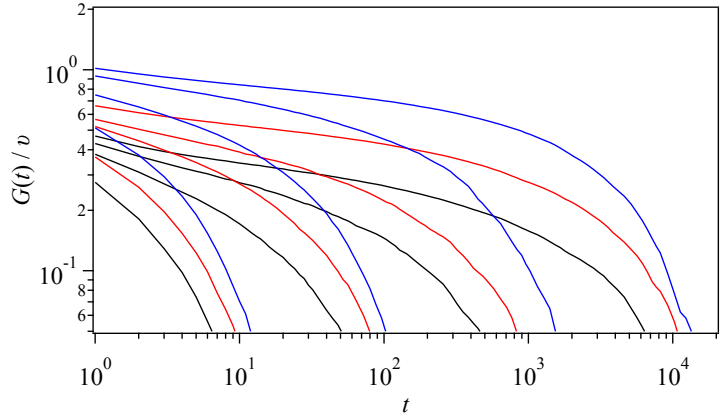

**Figure 3** Linear relaxation modulus $G(t)$ normalized by the segment density $\nu$, for $Z = 5, 10, 20, 40$ from left to right, with $K = 0$ (black), 0.3 (red), and 0.6 (blue).

Before discussing the link between $G_0$ and $L_p$, it is important to first consider the molecular weight dependence of $G_0$. Figure 4 shows $G_0$ plotted against $Z$ for various $K$. As reported earlier[14], $G_0$ increases with increasing $Z$, approaching an asymptotic value in the large-$Z$ limit. The following simple equation, proposed earlier[14], reasonably captures this $Z$-dependence of $G_0$

$$G_0 = G_\infty \left(1 - \frac{3}{Z}\right) \qquad (12)$$

where $G_\infty$ is the asymptotic value of $G_0$ in the large-$Z$ limit. Figure 4 shows that eq 12 works well also in the case of semiflexible chains. The dependence of $G_0$ on $Z$ has often been ignored in the past, also for the case of small-$Z$ systems. Here, we will refer to $G_\infty$, rather than to a $Z$-dependent $G_0$, to link modulus and rigidity.

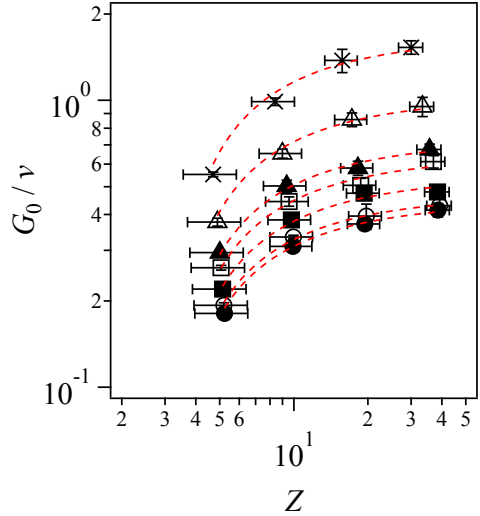

**Figure 4** Plateau modulus $G_0$ (normalized by the strand density $\nu$) plotted against the average strand number per chain $Z$, for various bending rigidities. $K$ values are 0 (filled circles), 0.1 (unfilled circles), 0.2 (filled squares), 0.3 (unfilled squares), 0.4 (filled triangles), 0.6 (unfilled triangles), and 1.0 (crosses). Horizontal error bars show $Z$ distributions, and vertical ones give standard deviations for eight independent simulations. Red dashed lines represent eq 12.

Finally, Figure 5(a) illustrates the desired relationship between the plateau modulus $G_\infty$ and the persistence length $L_p$. It is apparent that $G_\infty$ increases with increasing $L_p$, consistently with experiments. When fitting the results to a power-law function, the exponent is found to be close to 2/3, as shown by the red dashed line in Figure 5(a).

According to the Morse[3,4] theory, the relaxation modulus is a sum of different contributions, which were analyzed separately. Here, since the relaxation modulus is obtained from the stress autocorrelation, we can separate contributions to the stress tensor due to tensile (eq 3) and bending (eq 4) forces. The plateau moduli so obtained, $G_{0\_tens}$ and $G_{0\_bend}$, are shown in Figs 5(b) and 5(c), respectively, as a function of the average strand number per chain. Figure 5(b) indicates that $G_{0\_tens}$ is very close to the overall plateau modulus $G_0$ shown in Fig 4. Conversely, $G_{0\_bend}$ is significantly smaller. Since the tension along the chain is not modified by the bending potential, while, as expected, rigidity strongly affects the end-to-end distance (see Fig 2a), this result demonstrates that within our model the increase in modulus with increasing rigidity is not directly linked to the bending forces, but it is essentially due to chain swelling. Indeed, the stress is computed through the well-known Kramer's formula, i.e., as the dyadic product of force to position vector, and while the former is not significantly affected by rigidity, the latter increases a lot with increasing chain stiffness. The dependence of $G_{0\_tens}$ on $Z$ is well described by eq 12, as shown by the broken curves in Figure 5(b), whereas $G_{0\_bend}$ shows deviations for small-$Z$ values (see Figure 5c) . Nevertheless, the molecular weight dependence is not negligible even for these decomposed moduli.

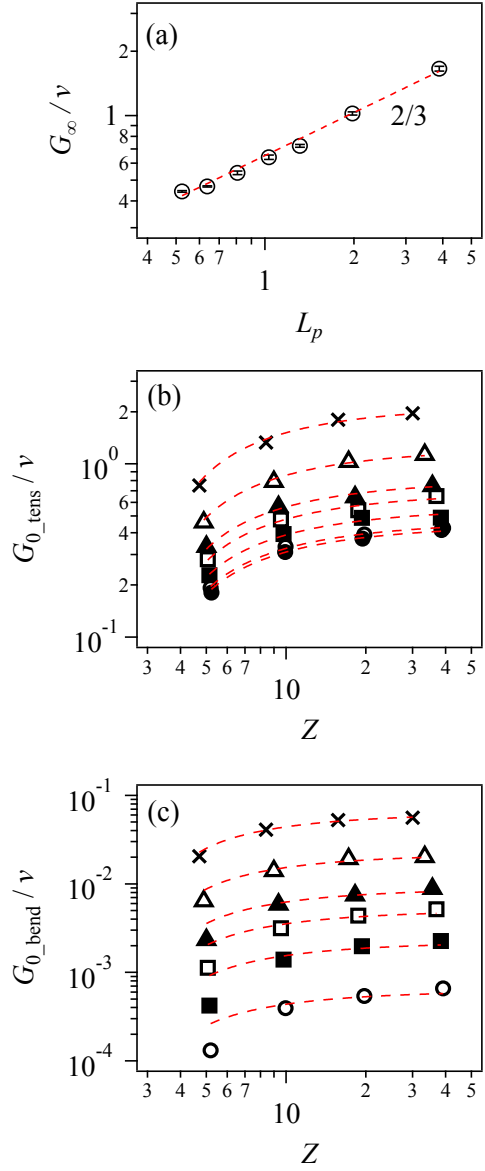

**Figure 5** (a) Asymptotic plateau modulus $G_\infty$ as a function of persistence length $L_p$; (b) contribution $G_{0\_tens}$ to the plateau modulus due to chain tension as a function of the average strand number per chain $Z$, and for various bending rigidity $K$; (c) similar plot for the contribution $G_{0\_bend}$ of the bending forces. In panel (a), the red broken line shows a slope of 2/3. In panels (b) and (c), the $K$ values are 0 (filled circles), 0.1 (unfilled circles), 0.2 (filled squares), 0.3 (unfilled squares), 0.4 (filled triangles), 0.6 (unfilled triangles), and 1.0 (crosses). Red broken curves are best fit from eq 12.

The relationship $G_\infty \propto L_p^{2/3}$ coming out from our simulations is not significantly different from experimental results for DNA solutions. Figure 6 shows the data reported by Schuldt et al.[9] for various DNA concentrations. Those authors reported that the slope of modulus vs persistence length is unity, shown by the red broken line in Figure 6. In the same figure, we also report our predicted slope of 2/3, indicated by the red solid line, which is also consistent with the data. It is fair to mention, however, that our prediction fails for the F-actin solutions, for which the reported exponent is close to five.[10,11]

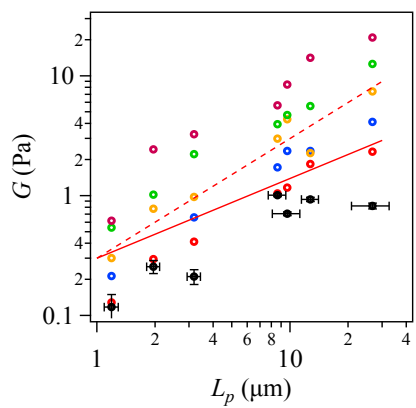

**Figure 6** The relationship between modulus and persistence length for DNA nanotube solutions reported by Schuldt et al.[9]. The concentration (in µM) is 13 (violet), 8 (green), 6 (orange), 4 (blue), 3 (red), and 2 (black), respectively. Red solid and broken lines indicate the slope of 2/3 and 1, respectively.

CONCLUSIONS

We extended the PCN model to semiflexible polymers by introducing bending rigidity of the primitive chain while (by necessity) ignoring rigidity in the chain strands between consecutive entanglements. The latter simplification is dictated by the coarse graining nature of the PCN model. We then performed equilibrium simulations (with varying molecular weights and bending rigidity) with the aim of calculating static and dynamic properties in the linear limit. The molecular weight dependence of the square end-to-end distance $R^2$ of the chains is found to follow the well-established relationship for worm-like semiflexible chains, showing the expected chain swelling due to rigidity. We then determined the persistence length $L_p$ by the same relationship linking $R^2$ to the contour length $L$ of the primitive chain. As regards dynamic properties, we calculated the linear relaxation modulus by means of the Green-Kubo formula and obtained the plateau modulus $G_0$. The molecular weight dependence of $G_0$ is found to follow the same behavior of flexible polymers, approaching an asymptotic value $G_\infty$ in the long chain limit. For $L_p$ and $G_\infty$ thus obtained we found the scaling $G_0 \propto L_p^{2/3}$, implying that $G_0$ increases with increasing $L_p$. This result is qualitatively consistent with experimental findings, though it contradicts some earlier theories. It is important to note, however, that in the PCN model, the entanglement length remains fixed regardless of rigidity, as we coarse grain at that level (with the entanglement length set to unity). Thus, the model would benefit from incorporating the relationship between bending and entanglement length. Studies are in progress to address this aspect, and the results will be reported elsewhere.

ACKNOWLEDGEMENTS

This study is partly supported by JST-CREST (JPMJCR1992) and JSPS KAKENHI (22H01189).